\documentclass[prl,twocolumn,showpacs,showkeys,amsfonts,amssymb]{revtex4}
\usepackage{amsmath}
\usepackage{graphics}
\usepackage{epstopdf}
\usepackage{graphicx}
\usepackage{color}

\begin{document}
\title{Force-Extension for DNA in a Nanoslit: Using an Effective Dimensionality to Map between the 3D and 2D Limits}
 \author{Hendrick W. de~Haan}
 \email{hendrick.dehaan@uoit.ca}
 \affiliation{Faculty of Science, University of Ontario Institute of Technology, 2000 Simcoe St. North, Oshawa, ON, L1H 7K4, Canada}
 \author{Tyler N. Shendruk}
 \affiliation{The Rudolf Peierls Centre for Theoretical Physics, Department of Physics, Theoretical Physics, University of Oxford, 1 Keble Road, Oxford, OX1 3NP, United Kingdom}
\begin{abstract}
The force-extension relation for a semi-flexible polymer such as DNA confined in a nanoslit is investigated and it is found that both the effective persistence length and the form of the force-extension relation change as the chain goes from 3D (very large slit heights) to 2D (very tight confinement).
Generalizations of the Marko-Siggia relation appropriate for polymers in nanoconfinement are presented.
The forms for both strong and weak confinement regimes are characterized by an \textit{effective dimensionality}.
At low forces, the effective dimensionality is given by the correlations along the polymer in the plane of the confining walls.
At high forces, the theoretical force must account for reduced conformation space.
Together the interpolations give good agreement for all slit heights at all forces.
As DNA and other semi-flexible biopolymers are commonly confined \textit{in situ} to various degrees, 
both the idea of an effective dimensionality and the associated generalized Marko-Siggia interpolations are useful for qualitatively understanding and quantitatively modeling polymers in nanoconfinement.

\end{abstract}
\maketitle

\section{Introduction}

The ability to confine single biopolymers such as DNA within nanoscale devices has yielded a wealth of research concerning both 
fundamental studies of the physics of polymers in confinement and the use of such devices for biomedical applications~\cite{reisner12}. 
Many recent studies have focused on the static~\cite{lin07,dimitrov08,dekker08,cifra09,tang10a,dai12a,cifra12,hsu13,chen14} and dynamic~\cite{hsieh07,strychalski08,tang10b,michev11,trahan11,dai13b} properties of semiflexible polymers within nanoslits as well as nanochannels ~\cite{reisner05,wang11,tree12,tree13,tree13b}. 

Here we study polymers confined in a nanoslit and subject to a stretching force, as in tug-of-war and nanopit-type devices~\cite{reisner09,stein09,klotz12,yeh12,shager13}. 
We focus on the relationship between the force $F$ applied to the ends of the polymer and the resulting extension $X$.
The force extension relation has been studied extensively in both the 3D~\cite{bustamante00,bustamante03} and 2D~\cite{hsu11} limits. 
Few studies have investigated the transition between the two limits as a function of confinement~\cite{taloni13}. 

Chen \textit{et al.} examined this transition via computer simulations and compared it to a force-extension relation that was proposed without derivation~\cite{chen10}. 
In that work, the extension was defined as the absolute value between the ends of the polymer in the direction of the force rather than a vector quantity.
Consequently, the extension did not go to zero with force and the low-force regime was unresolved. 
In many nanofluidic applications, the low-force regime that is of significant interest and, herein, we obtain the force-extension curve for both the low and high-force regimes.
We use an effective dimensionality $d_\textmd{eff}$ to map between the 2D to 3D limits. 
It is found that $d_\textmd{eff}$ depends not only on the slit height, but also on the applied force $F$.
Taking both effects into consideration, we predict the simulated force-extension curves for all slit heights and both force regimes. 

The response of a linear semi-flexible polymer of contour length $L_\textmd{C}$ to a force pulling on each end can be separated into two regimes. 
At low stretching forces, the polymer behaves as an entropic spring. 
The low-force linear relation $\tilde{x} \equiv \tfrac{X}{L_\textmd{C}}=\tfrac{2}{d} \tfrac{F L_{\xi}}{k_\textmd{B}T}$ can be derived from a Kratky-Porod worm-like chain model~\cite{kratkyPorod49} in dimensionality $d$. 
The low-force limit uses the chain's persistence length $L_{\xi}$, which defines the correlation length between tangent vectors along the polymer. 

In the strong-force limit in $d$-dimensions, the equipartition theorem can be used in Fourier space to find that $\tilde{x}=1-\Delta \tilde{f}_{\kappa}^{-1/2}$, where the dimensionless force $\tilde{f}_{\kappa}\equiv F L_{\kappa}/k_{B}T$ required to stretch a polymer diverges as $\tilde{x} \rightarrow 1$~\cite{prasad05} and $\Delta=\left(d-1\right)/4$ is $\Delta=1/2$ in 3D and $1/4$ in 2D. 
The intrinsic length scale of mechanical rigidity $L_{\kappa} = \kappa/k_{B}T$ enters this limit through the energetic cost of bending. 
While the rigidity $\kappa$ is a material property of the DNA, the persistence length $L_{\xi}$ is a statistical value that depends on dimensionality. 
In this work, the force is always nondimensionalized using $L_{\kappa}$ as emphasized by the subscript. 
Interpolating between the two limits produces a generalized Marko and Siggia relation for discrete dimensions:
\begin{align}
  \frac{\tilde{f}_{\kappa}\left(\tilde{x},d,L_{\xi}\right)}{\Delta^{2}} &=  \left(1-\tilde{x}\right)^{-2}-1+\left(\frac{8d}{\Delta^2}\frac{L_{\kappa}}{L_{\xi}}-2\right)\tilde{x} \label{MS_CORR},
\end{align}
which is consistent with the Marko-Siggia form in 3D~\cite{markoSiggia95} and 2D~\cite{prasad05}. 

\begin{figure}[tb]
 	\centering
	\includegraphics[width=0.50\textwidth]{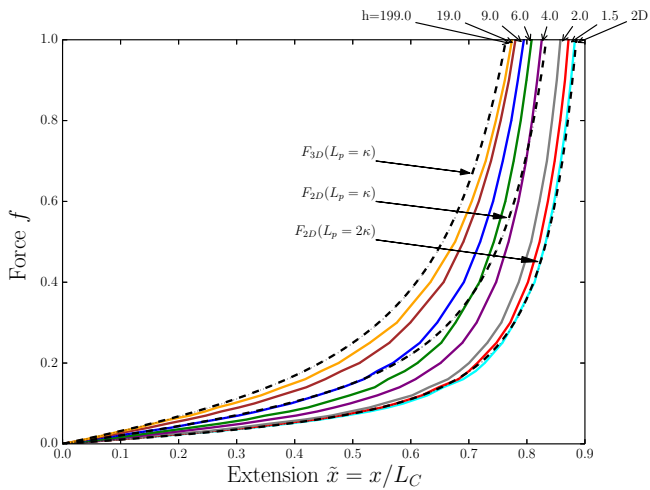} 
	\caption{Force-extension relation as found form the simulations for various slit heights.
	Dashed lines indicate theoretical curves corresponding to the 2D and 3D limits of Eqn. \ref{MS_CORR}.
	In 2D, the curves using both $L_{\xi} = L_{\kappa}$ and $L_{\xi} = 2L_{\kappa}$ are shown.
	}
	\label{fig:full}
\end{figure}

For the computational polymer model, neighbouring monomers are joined via a FENE potential, while overlap is prevented via a WCA potential~\cite{slater2009}. 
All lengths are given in units of the WCA lengthscale $\sigma$ and energy in $k_{B}T$. 
Standard values are used for the constants in the forces, which results in a bond length of $b \approx 0.97$. 
Polymer rigidity is implemented via a harmonic bending potential between monomers with a spring constant $k=5.0$. 
The chain consists of 200 monomers ($L_\textmd{C} = 200 b$).
To ease comparison to theory, simulations are performed with an ideal polymer and thus there are no interactions between non-neighbouring monomers.
The interaction between monomers and the confining walls is given by the WCA potential.
After equilibration during which a force $F$ is applied in opposite directions to each end of the polymer, the average extension is measured. 
This process is repeated for 10 independent runs in order to adequately resolve the average extension.
The results for different slit heights are shown in Fig. \ref{fig:full}. 
Note that polymers of size $N = 300, 400$ were simulated at $h=19.0$ but no appreciable difference in the force-extension curves was observed. 

In Fig. \ref{fig:full}, reasonable agreement between simulations in 3D (indistinguishable from the largest slit height shown, $h=199$) and Eqn. \ref{MS_CORR} is obtained for $d=3$ and $L_{\xi}=L_{\kappa}$, though theory slightly overpredicts the simulations.
This shift to larger extensions in the 3D limit is a limitation of the simulations. 
The dependence of persistence length on the dimensionality and rigidity length is known:
\begin{align}
L_{\xi}\left(L_{\kappa},d\right) &= \frac{L_{\kappa}}{2\Delta} = \frac{2L_{\kappa}}{d-1}. 
\label{L_corr}
\end{align}
However, calculating $L_{\xi}$ for different values of harmonic spring constant $k$ demonstrates a small deviation at finite $k$. 
We calculate $L_{\xi} = 5.177$ for $k=5.0$, which differs by only 3.5\% from Eqn. \ref{L_corr}. 
We thus approximate $L_{\xi}\approx5.0$ in 3D.

The 2D force-extension curve is shifted to larger extensions at equivalent forces compared to the 3D curve. 
The smallest slit heights approach the purely 2D simulations. 
Eqn. \ref{MS_CORR} fails to agree with the 2D simulations if the rigidity lengthscale $L_{\kappa}$ is erroneously utilized as the persistence length as was done in 3D --- not only do the coefficients of the Marko-Siggia relation change but the persistence length does as well. 
From Eqn. \ref{L_corr}, $L_{\xi} = 2 L_{\kappa}$ in the 2D limit.

Substituting $L_{\xi}$ from Eqn. \ref{L_corr} into  Eqn. \ref{MS_CORR} produces the generalized Marko-Siggia relation without reference to correlation length to be
\begin{align}
  \frac{\tilde{f}_{\kappa}\left(\tilde{x},d\right)}{\Delta^{2}}  &= \left(1-\tilde{x}\right)^{-2}-1+\left(\frac{d+1}{2\Delta}\right)\tilde{x} \label{MS_DIM},
\end{align}
which is in excellent agreement with both the 3D and 2D limits. 
Eqn. \ref{L_corr} is seen to hold to higher precision for $k \gtrapprox 2$ in the 2D limit than in 3D. 

Equation \ref{MS_DIM} represents a unified form for the previously-known cases of discrete dimensions, which avoids persistence length by utilizing only the intrinsic rigidity lengthscale and dimensionality $d$. 
However, it is the transition from the 3D to 2D limit that interests us. 

\begin{figure}
 	\centering
	\includegraphics[width=0.50\textwidth]{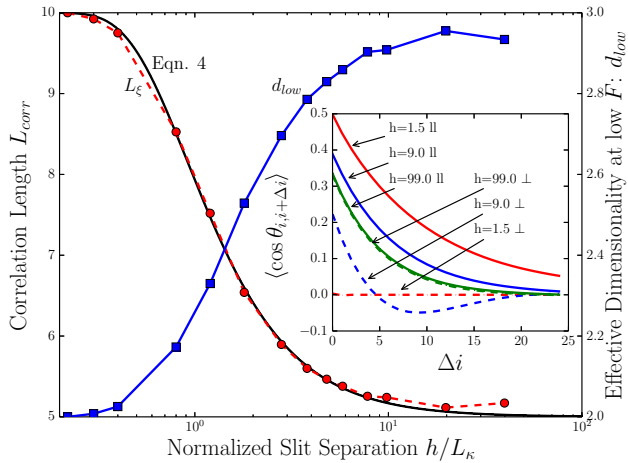} 
	\caption{Correlation length (dashed circles) and corresponding effective dimensionality (solid squares) as a function of the slit height.
	The solid black lines is a fit to the data for $h \leq 2L_{\kappa}$ (Eqn. \ref{fit}).
	The inset shows the correlations of direction vectors along the polymer contour as a function of the distance between the vectors.
	Both the parallel (in-plane) and perpendicular components are shown. 
	}
	\label{fig:Lp}
\end{figure}

We use Eqn. \ref{L_corr} to define an effective dimensionality $d_\textmd{low}=1+2L_{\kappa}/L_{\xi}$ as a function of correlation length measured at particular finite slit heights to map out this intermediate slit behaviour from the 3D to 2D extremes. 
The correlation length is typically measured via correlations between angle vectors, $\langle \cos \theta_{i, i+\delta i} \rangle$, for polymer segments of increasing distance $\delta i$ using $\langle \cos \theta_{i, i+\delta i} \rangle \equiv e^{- \delta i / L_{\xi}}$. 
While this approach works well in the 2D and 3D limits, the results for intermediate heights do not conform to a single exponential decay since the correlations are non-isotropic, as seen in Fig. \ref{fig:Lp} (inset).
At intermediate heights, the parallel components remain a smooth decay, but the perpendicular component becomes non-monotonic
as the correlations become negative for intermediate separations. 
This anti-correlation arises from the reflections of the polymer off of the slit walls in the $z$ direction and so $L_{\xi}$ is not simply defined. 
Fully understanding the correlation functions of semiflexible polymers in confinement remains challenging experimentally~\cite{koster05,koster07,koster08,noding12}, computationally~\cite{cifra08,cifra08b,benkova12}, and analytically~\cite{harnau99,choi05,wagner07}. 

We follow the approach of Chen \textit{et al.}~\cite{chen10} and use the parallel correlation measurements to define $L_{\xi}\left(h\right)$ as shown in Fig. \ref{fig:Lp}. 
As expected, $L_{\xi}$ transitions from the 3D value to the 2D value with decreasing slit height and, correspondingly, the effective dimensionality $d_\textmd{low}\left(L_{\xi}\right)$ defined via Eqn. \ref{L_corr} varies smoothly from about $3$ to $2$. 
The black solid line is a fit of the data for $h \leq 2 L_{\kappa}$ given by 
\begin{align}
 L_{\xi}\left(L_{\kappa},h\right) = L_{\kappa} \left[ 2 - e^{-0.88 \left( L_{\kappa}/h \right)^{1.41}} \right],
 \label{fit}
\end{align}
which is similar in form to that given by Chen \textit{et al.}~\cite{chen10}., but agrees with theory for both $h \rightarrow 0$ and $h \rightarrow \infty$. 

We previously gave Eqn. \ref{MS_DIM} to be a generalized form of the Marko-Siggia relation for arbitrary discrete dimensions $d$ without explicit reference to the correlation length $L_{\xi}$. 
We propose that the concept of effective dimensionality $d_\textmd{low}$ can be substituted into Eqn. \ref{MS_DIM} in place of the discrete dimensionality $d$. 
For now, we assume that dimensionality is only a function of correlation length $d_\textmd{low}\left(L_{\xi}\right)$. 
Doing so allows the Marko-Siggia relationship given by Eqn. \ref{MS_DIM} to apply to finite slit confinements as a function of measured correlation length. 

\begin{figure}
 	\centering
	\includegraphics[width=0.50\textwidth]{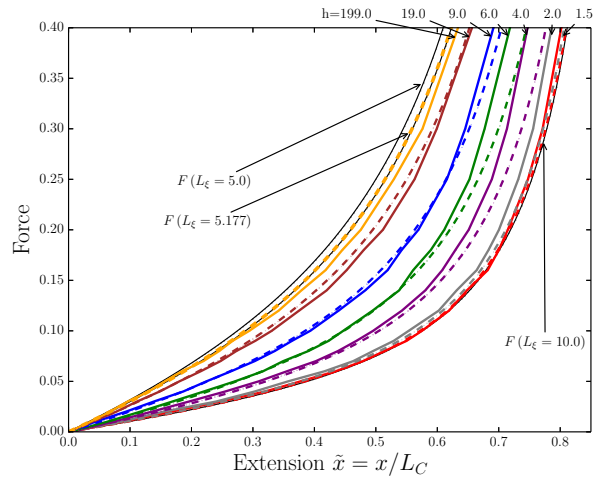} 
	\caption{Force-extension curves as measured by simulations (solid curves) and predicted by Eqn. \ref{MS_DIM} (dashed lines) for the low-force regime and various slit heights.}
	\label{fig:low}
\end{figure}

The results are shown in Fig. \ref{fig:low}.
Beyond the previously mentioned deviation in the 3D limit, good agreement between theory and simulations is obtained for low forces, corresponding to low extensions ($ \tilde{x} < 0.5$).
Hence, the effective dimensionality approach maps out the transition from 3D to 2D as a function of slit height at low forces.
However, the agreement is lost as intermediate forces and large extensions are reached. 
The effective-dimensionality theory over-predicts the extension in comparison to the Langevin simulations. 
This suggests that expressing the extension in terms of effective dimensionality as is only accurate in the low-force regime
\begin{align}
 \lim_{\tilde{f}_{\kappa} \ll 1} \tilde{x} &= \frac{4\tilde{f}_{\kappa}}{d_\textmd{low}\left(d_\textmd{low}-1\right)}. 
 \label{low}
\end{align}

Consider a confined polymer that is nearly fully extended (\textit{i.e.}, $\tilde{x} \rightarrow 1$). 
Such a taut polymer can only accommodate small thermal fluctuations about the line connecting the end monomers and so does not feel the effect of the walls. 
In other words, strong forces alone limit the conformations available to the polymer and confinement plays a diminished role. 
Hence, the confinement effects diminish as the force increases causing the effective dimensionality to increase and thus to also depend on the applied force. 
By applying the equipartition theorem to the energy of small thermal deflections in Fourier space with a cutoff frequency due to the confining walls, we find the expression for the extension in the strong force limit and arbitrary confinement to be 
\begin{small}
\begin{align}
 \lim_{\tilde{f}_{\kappa} \gg 1} \tilde{x} &= 1-\frac{\Delta}{\tilde{f}_{\kappa}^{1/2}} \left[1-\frac{1}{2\pi\Delta} \sum_{i=0}^{d^{\prime}} \tan^{-1}\left(\frac{c_0}{\tilde{f}_{\kappa}^{1/2}\frac{L_{\kappa}}{h_i}}\right)\right],
 \label{beam}
\end{align}
\end{small}
where $c_0$ controls the cutoff frequency and the summation is only over the $d^{\prime}$ confined dimensions. 
For a slit, $d=3$, $d^{\prime}=1$. 
In this derivation, the polymer has no finite width and thus to compare between simulations and theory, we set $h_{theory} = h_{sim}-1$.
Good agreement is found in the high-force limit for all slit heights (not shown). 

The arctangent in Eqn. \ref{beam} hinders the development of a generalized interpolation that is accurate in both limits. 
Hence we seek to interpolate between Eqn. \ref{low} and the Taylor expansions of Eqn. \ref{beam} when confinement dominates over force or when the force dominates. 
For this reason, we must consider the argument of the arctangent, which is the competition of confinement effects $L_{\kappa}/h$ against the force $\tilde{f}_{\kappa}^{1/2}$. 
Interpolation can be found in either limit but not for the general strong-force regime. 

\begin{figure}[tb]
 	\centering
	\includegraphics[width=0.50\textwidth]{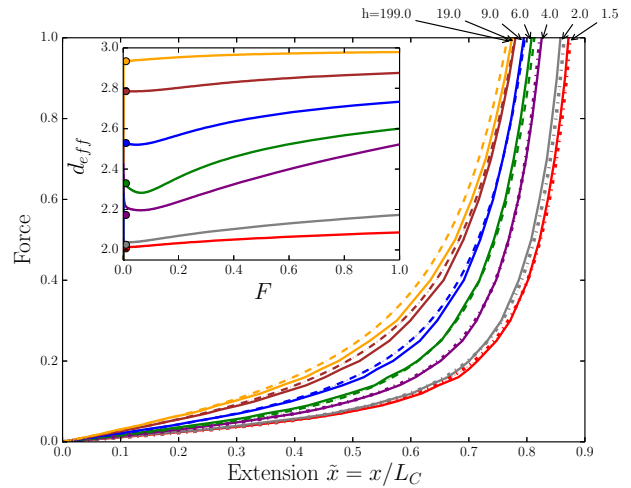} 
	\caption{The force extension curves measured in simulations (solid lines) and Eqn. \ref{forceDominated} for $h>L_{\kappa}$ (dashed lines) or Eqn. \ref{confinementDominated} for $h<L_{\kappa}$ (dash-dot lines). 
	The inset shows the dependence of effective dimensionality $d_\textmd{eff}$ on the force for different slit heights for different slit heights.
	The points at $F=0$ indicate the $d_{low}$ values as calculated form the in-plan correlations.}
	\label{fig:high}
\end{figure}

Let us consider the force-dominated limit of the strong-force regime ($\tilde{f}_{\kappa}^{1/2} \gg L_{\kappa}/h$). Interpolation with the low-force limit gives
\begin{align}
 \frac{\tilde{f}_{\kappa}}{\Delta^2 } &=  \left(1-\tilde{x}\right)^{-2} - \frac{2A}{\Delta^2}\left(1-\tilde{x}\right)^{-1} - \left(1-\frac{2A}{\Delta^2}\right) \nonumber\\
 &\qquad + 2\left\{ \frac{d_\textmd{low}\left(d_\textmd{low}-1\right)}{8\Delta^2} - \left(1-\frac{A}{\Delta^2}\right)\right\} \tilde{x} ,
 \label{forceDominated}
\end{align}
which depends on both the natural dimensionality $d$ through $\Delta$ and the effective low-limit dimensionality $d_\textmd{low}$, as well as the confinement through $A = c_0 \sum_{i=0}^{d^{\prime}} L_{\kappa} / h_i$. 
For a 3D slit, $A=c_0 L_{\kappa} / h$ where $c_0=0.3$ is found to give good agreement. 
Fig. \ref{fig:high} shows that Eqn. \ref{forceDominated} is highly accurate at both low and high forces when the slit height is large ($h \gtrsim L_{\kappa}$). 
However, the weak confinement approximation leading to Eqn. \ref{forceDominated} breaks down as the slit height decreases and the effective dimensionality moves towards to 2. 

The other limit of Eqn. \ref{beam} is confinement dominated. 
Interpolating the confinement dominated limit with the low force limit produces
\begin{small}
\begin{align}
 \frac{\tilde{f}_{\kappa}}{{\Delta^{\prime}}^2} &= \left(1-\left\{\tilde{x}+B\right\}\right)^{-2} - \left(1+2B\right) + \left(\frac{d_\textmd{low}\left(d_\textmd{low}-1\right)}{4{\Delta^{\prime}}^2}-2\right)\tilde{x}
 \label{confinementDominated}
\end{align}
\end{small}
where $\Delta^{\prime}=\left(d-d^{\prime}-1\right)/4$, $B=\left(\sum_{i=0}^{d^{\prime}} h_i/L_{\kappa}\right)/2\pi c_0$, and $c_0$ is set to 1.55 to obtain good agreement. 
In a slit, $\Delta^{\prime}=1/4$ and $B=h/\left(2\pi c_0 L_{\kappa}\right)$. 
Figure \ref{fig:high} demonstrates that this interpolation is accurate for small slit heights ($h \lesssim L_{\kappa}$). 
While Eqn. \ref{forceDominated} represented a correction on the 3D form of Eqn. \ref{MS_DIM} due to confinement, Eqn. \ref{confinementDominated} represents a correction on the 2D form.  
In this work, the interpolations are applied to slits but are also predicted to hold in asymmetrical channels. 
However, if one were to consider a confinement-dominated channel, rather than slit, then $\Delta^{\prime}=0$ and the low force limit would breakdown since the chain is nearly fully-extended even in this limit. 

Having generalized Marko-Siggia interpolations for force extension in a slit, we return to the concept of an effective dimensionality in Eqn. \ref{MS_DIM}, but now recognize that effective dimensionality $d_\textmd{eff}$ is a function of both correlation length and force. 
We extract $d_\textmd{eff}\left(F,h\right)$ by fitting Eqns. \ref{forceDominated} and \ref{confinementDominated} to the generalized Marko-Siggia equation (Eqn. \ref{MS_DIM}). 
The resulting effective dimensionality is shown in Fig. \ref{fig:high} (inset). 
Apart from some spurious behaviour at very low forces near the strong to weak confinement transition, reasonable results are obtained across $h$ and $F$: 
the curves start at $d_\textmd{eff}=d_\textmd{low}$ --- which increases as $h$ increases --- and increase as $F$ increases.
Further, the drift towards 3D is slower at small $h$ than for intermediate heights;
\textit{i.e.}, confinement effects continue to have an impact even for relatively large stretching forces at very tight confinement.

We have presented a physical picture of the force-extension relation for DNA confined within a nanoslit by introducing an effective dimensionality, $d_\textmd{eff}$.
Using $d_\textmd{eff}$ in a generalized Marko-Siggia relation leads to good agreement with the simulation data for all slit heights at low and high forces.
At low forces, the effective dimensionality is determined from the in-plane, parallel correlations of the direction vectors along the polymer.
However, as the force increases, the effect of confinement decays and the effective dimensionality drifts towards 3.
Via interpolation, we derived force-extension relations for force-dominated (near 3D) and confinement-dominated (near 2D) systems.
These semi-empirical formulas were shown to give good agreement with simulation results. 
Comparison to the generalized Marko-Siggia yields $d_\textmd{eff}$ as a function of both slit height and stretching force.
A $d_\textmd{eff}$ that starts from values closer to 2 for tighter confinement but tends towards 3 at all slit heights thus serving as a useful physical picture for the relatively complicated nature of the force-extension curve for polymers in confinement.
Future research should consider to what extent the concept of effective dimensionality could more generally be applied to confined DNA~\cite{huang14}. 

Simulations were performed using the HOOMD Blue simulation package~\cite{anderson08} on the SHARCNET computer system (www.sharcnet.ca).

\bibliographystyle{apsrev4-1}
\bibliography{wlcSources.bib}

\end{document}